\documentclass[prd,twocolumn,groupedaddress,noshowpacs,nofootinbib]{revtex4}
\usepackage{graphicx}
\usepackage{dcolumn}
\usepackage{amssymb}
\usepackage{mathrsfs}
\usepackage{amsmath}
\usepackage{epsfig}
\usepackage[dvips]{color}
\usepackage{hhline}

\begin{document}

\title{	Efficient approach to low scale Peccei-Quinn symmetry breaking without domain wall problem}

\author{Pei-Hong Gu}

\email{phgu@seu.edu.cn}

\affiliation{School of Physics, Jiulonghu Campus, Southeast University, Nanjing 211189, China}

\begin{abstract}

We propose an efficient mechanism to realize an invisible axion from a low scale Peccei-Quinn symmetry breaking. Our basic model only contains a gauge boson, an up-type vector-like quark, two Higgs doublets and two Higgs singlets besides the standard model fermions and gauge bosons. The physical Peccei-Quinn global symmetry is a result of two independent global symmetries connected by the new gauge symmetry. Anyone of these two global symmetries only acts on either the right-handed top quark or the left-handed new quark so that it can avoid the domain wall problem. Thanks to the electroweak and new gauge interactions, the Higgs doublet for the top quark mass generation and the Higgs singlet for the new quark mass generation can only contribute a tiny fraction in the axion. The axion decay constant can be largely enhanced by a factor composed of the vacuum expectation values of the four Higgs scalars.

\end{abstract}


\maketitle

\section{Introduction}

Currently the global Peccei-Quinn (PQ) symmetry \cite{pq1977} $U(1)_{PQ}^{}$ has become the most popular scheme to solve the so-called strong CP problem. In this scenario, a Nambu-Goldstone (NG) boson from the spontaneous PQ symmetry breaking picks up a mass through the color anomaly \cite{adler1969,bj1969,ab1969}, and hence becomes a pseudo NG boson, named the axion \cite{weinberg1978,wilczek1978}. Since the axion has not been observed in any experiments, it can only interact with the standard model (SM) particles at an extremely weak level \cite{kg2010,dgnv2020,pdg2020}. For this purpose, the Kim-Shifman-Vainstein-Zakharov (KSVZ) \cite{kim1979,svz1980} and Dine-Fischler-Srednicki-Zhitnitsky (DFSZ) \cite{dfs1981,zhitnitsky1980} models were proposed  after the original PQ model was quickly ruled out. In the KSVZ and DFSZ models \cite{kim1979,svz1980,dfs1981,zhitnitsky1980}, the PQ symmetry should be spontaneously broken far far above the weak scale to make the axion invisible \cite{kg2010,dgnv2020,pdg2020}.

In the invisible axion models, the accompanying new particles would be too heavy, i.e. of the order of the PQ symmetry breaking scale, to leave any experimental signals, unless the related couplings were artificially small. In other words, all experimental attempts to discover the axion can only depend on the mixing of the axion to the $\pi$ and $\eta$ mesons \cite{kg2010,dgnv2020,pdg2020}. A high scale PQ symmetry breaking also has some theoretical disadvantages. For example, the more high the PQ symmetry breaking scale is, the more serious the axion quality problem becomes \cite{dgnv2020}. Moreover, for a huge hierarchy between the PQ and electroweak symmetry breaking scales, unless the inevitable Higgs portal has an extremely small coupling, its contribution should have a large cancellation with the rarely quadratic term of the SM Higgs scalar \cite{cv2016}. In order to realize a low scale PQ symmetry breaking, people have tried to consider an anomalous gauge symmetry \cite{mrs2001} or enormous scalar fields \cite{cky2014,kr2016,ci2016,gu2021}.

It was pointed out \cite{sikivie1982} after the $U(1)_{PQ}^{}$ global symmetry was explicitly broken by the QCD gluon anomaly, its discrete subgroup $Z_{N_{DW}^{}}^{}$ could remain unbroken and hence there would be $N_{DW}^{}$ vacua with exactly same physical properties. Here the index $N_{DW}^{}$ is nothing but the number of the quarks transform under the $U(1)_{PQ}^{}$ symmetry. The early universe thus could suffer a serious problem of domain wall formation after the spontaneous breaking of the $U(1)_{PQ}^{}$ global symmetry and then the $Z_{N_{DW}^{}}^{}$ discrete symmetry. One solution to this problem is to render a unique vacuum by assuming that only one quark is charged under the $U(1)_{PQ}^{}$ symmetry \cite{pwy1986,kw1986}, like the original KSVZ model just with one vector-like quark. However, it seems nontrivial to do such an assignment in the DSFZ-like models with two or more Higgs doublets, where an up-type quark must be accompanied by a down-type quark to simultaneously couple with the axion \cite{dfs1981,zhitnitsky1980}.

On the other hand, the top quark is the unique SM fermion whose Yukawa coupling to the SM Higgs boson is of the order of unity. Being heavier than a $W^\pm_{}$ boson, it is the only quark to decay before hadronization and have a very short lifetime. Therefore, the top quark plays a special role in the SM \cite{pdg2020}. This special role implies that the top quark may have a different mass origin from the other SM fermions. The heavy top quark mass can be argued by a proper PQ symmetry, under which the right-handed top quark is the unique SM fermion carrying a nontrivial PQ charge \cite{pwy1986,kw1986,cfty2015,cfty2017,ctty2018} so that it can exclusively couple to one Higgs doublet with a bigger vacuum expectation value (VEV) while the other SM fermions can only couple to another Higgs doublet with a smaller VEV.

\begin{table*}
\vspace{0.25cm}
\begin{center}
\begin{tabular}{|l|c|c|c|c|c|c|c|c|c|c|}  \hline &&&&&&&&&&\\[-2.0mm] ~~$\textrm{The scalars and fermions}$~~~~~~&&~~$\xi$~~ &~~$\sigma$~~&~~$\phi$~~&~~ $\eta$~~ &&~~$t_R^{}$~~ & ~~$T_L^{}$~~& ~~$T_R^{}$~~&  ~~$\textrm{The~other~SM~fermions}$~~ \\
&&&&&&&&&&\\[-2.0mm]
\hline&&&&&&&&&& \\[-1.5mm]~~The~$U(1)_{X}^{}$~gauge~symmetry~~ &&$+\frac{1}{2}$  & $+\frac{1}{2}$& $+\frac{1}{2}$ &$0$& & $+\frac{1}{2}$ & $+\frac{1}{2}$&$0$ &$0$ \\
&&&&&&&&&&\\[-2.0mm]
\hline&&&&&&&&&& \\[-1.5mm]~~The~$U(1)_{PQ}^{}$~global~symmetry~~ &&$-1$  & $x\neq - 1$& $-1$ & $0$ & & $-1$ & $x\neq - 1$&$0$ &$0$ \\[1.5mm]
\hline
\end{tabular}
\vspace{0.25cm}
\caption{\label{charge} The scalars and fermions in our basic models. The scalars $\xi$ and $\sigma$ are the Higgs singlets while $\phi$ and $\eta$ are the Higgs doublets. The new quarks $T_{L,R}^{}$ carry an electric charge $+\frac{2}{3}$ as same as the up-type quarks. The right-handed top quark is the unique SM fermion carrying the nonzero quantum numbers under the $U(1)_X^{}$ and $U(1)_{PQ}^{}$ symmetries. While the $U(1)_X^{}$ gauge symmetry is free of anomaly, the $U(1)_{PQ}^{}$ global symmetry keeps anomalous.}
\end{center}
\end{table*}

In the present work we shall propose an efficient mechanism to realize an invisible axion from a spontaneous PQ symmetry breaking at low scales. In our basic model, the right-handed top quark is the unique SM fermion charged under a new $U(1)_X^{}$ gauge symmetry. Then a new left-handed up-type quark and its right-handed partner are introduced to cancel the gauge anomalies. The new left-handed and right-handed quarks construct their Yukawa coupling with a Higgs singlet for spontaneously breaking the $U(1)_X^{}$ gauge symmetry. Another Higgs singlet dominates the spontaneous breaking of the $U(1)_{PQ}^{}$ global symmetry. There is also a Higgs doublet coupling to the SM fermions except for the top quark. A second Higgs doublet exclusively couples to the top quark and dominantly leads to the electroweak symmetry breaking. Because of the gauge interactions, the Higgs doublet for the top quark mass generation as well as the Higgs singlet for the new quark mass generation can only contribute a negligible fraction to the axion. The axion decay constant can be largely enhanced by a factor composed of the VEVs of the four Higgs scalars. In our mechanism, the $U(1)_{PQ}^{}$ symmetry indeed is a result of two independent $U(1)$ global symmetries connected by the $U(1)_X^{}$ gauge symmetry. Every $U(1)$ global symmetry only acts on either the right-handed top quark or the left-handed new quark. So, the spontaneous breaking of the two $U(1)$ global symmetries and hence the physical $U(1)_{PQ}^{}$ global symmetry can avoid the domain wall problem.

\section{Basic model}

We now begin to demonstrate our mechanism in details. The scalars and fermions in our basic model are summarised in Table \ref{charge}. The scalars $\xi$ and $\sigma$ are the Higgs singlets while $\phi$ and $\eta$ are the Higgs doublets. The new quarks $T_{L,R}^{}$ carry the electric charge $+\frac{2}{3}$ as same as the up-type quarks. The right-handed top quark is the unique SM fermion carrying the nonzero quantum numbers under the $U(1)_X^{}$ and $U(1)_{PQ}^{}$ symmetries. While the $U(1)_X^{}$ gauge symmetry is free of anomaly, the $U(1)_{PQ}^{}$ global symmetry keeps anomalous. The full scalar potential is 
\begin{eqnarray}
\label{potential}
V&=& \mu_\sigma^2 \sigma^\dagger_{} \sigma +  \mu_\xi^2\xi^\dagger_{}\xi  + \mu_\phi^2 \phi^\dagger_{}\phi +\mu_\eta^2 \eta^\dagger_{}\eta  +\lambda_\sigma^{} (\sigma^\dagger_{}\sigma)^2_{}\nonumber\\
&&+\lambda_{\xi}^{}(\xi^\dagger_{}\xi)^2_{}+\lambda_\phi^{} (\phi^\dagger_{}\phi )^2_{} +\lambda_\eta^{} 
(\eta^\dagger_{}\eta)^2_{}+\lambda_{\sigma\xi}^{} \sigma^\dagger_{}\sigma\xi^\dagger_{}\xi \nonumber\\
&& + \lambda_{\sigma\phi}^{} \sigma^\dagger_{}\sigma \phi^\dagger_{}\phi + \lambda_{\sigma\eta}^{}\sigma^\dagger_{}\sigma \eta^\dagger_{}\eta + \lambda_{\xi\phi}^{}\xi^\dagger_{}\xi \phi^\dagger_{}\phi \nonumber\\
&& +\lambda_{\xi\eta}^{}\xi^\dagger_{}\xi \eta^\dagger_{}\eta +\lambda_{\phi\eta}^{}  \phi^\dagger_{}\phi \eta^\dagger_{}\eta
+\lambda'^{}_{\phi\eta} \eta^\dagger_{}\phi \phi^\dagger_{}\eta \nonumber\\
&&+\rho (\xi\phi^\dagger_{}\eta+\textrm{H.c.})\,,
\end{eqnarray}
where the gauge invariant term $\lambda'^{}_{\sigma\xi}\left[\left(\sigma^\dagger_{}\xi\right)^2_{}+\textrm{H.c.}\right]$ has been forbidden by the $U(1)_{PQ}^{}$ global symmetry.  All of the allowed Yukawa interactions are 
\begin{eqnarray}
\label{yukawa}
\mathcal{L}_{Y}^{} &=& -y_{i e}^{} \bar{l}_{Li}^{} \eta e_R^{} -y_{i\mu}^{} \bar{l}_{Li}^{} \eta \mu_R^{}  -y_{i \tau}^{} \bar{l}_{Li}^{} \eta \tau_R^{} \nonumber\\
&&  -y_{i d}^{} \bar{q}_{Li}^{} \eta d_R^{}  - y_{is}^{} \bar{q}_{Li}^{} \eta s_R^{} - y_{ib}^{} \bar{q}_{Li}^{} \eta b_R^{}   \nonumber\\
&& -y_{i u}^{} \bar{q}_{Li}^{} \tilde{\eta} u_R^{} -y_{ic}^{} \bar{q}_{Li}^{} \tilde{\eta} c_R^{}- y_{iT}^{} \bar{q}_{Li}^{}\tilde{\eta}T_R^{}\nonumber\\
&&-y_{i t}^{} \bar{q}_{Li}^{} \tilde{\phi} t_R^{}  -y_{Tu}^{} \sigma \bar{T}_L^{} u_R^{}  -y_{Tc}^{} \sigma \bar{T}_L^{} c_R^{} \nonumber\\
&&-y_{TT}^{} \sigma \bar{T}_L^{} T_R^{} +\textrm{H.c.} \,.
\end{eqnarray}
Without loss of generality, two of the three Yukawa couplings $y_{Tu}^{}$, $y_{Tc}^{}$ and $y_{TT}^{}$ can be removed by a proper phase rotation. For convenience we can set $y_{Tu}^{}=y_{Tc}^{}=0$ and $y_{TT}^{}=y_T^{}$.

We emphasize that the above $U(1)_{PQ}^{}$ global symmetry indeed should be treated as two independent $U(1)$ global symmetries: a $U(1)_t^{}$ only acts on the right-handed top quark $t_R^{}$, the Higgs doublet $\phi$ and the Higgs singlet $\xi$, while a $U(1)_T^{}$ only acts on the left-handed new quark $T_L^{}$ and the Higgs singlet $\sigma$. Therefore, the $U(1)_t^{}$ and $U(1)_T^{}$ global symmetries will not suffer the domain wall problem. As shown later, only one of the two NG bosons from the $U(1)_t^{}$ and $U(1)_T^{}$ global symmetry breaking can survive from the gauge interactions to serve as the axion. This means the physical $U(1)_{PQ}^{}$ global symmetry should be a result of the $U(1)_t^{}$ and $U(1)_T^{}$ global symmetries.

\section{Physical scalars}

After the gauge and global symmetries are all spontaneously broken, i.e. $SU(2)_L^{}\times U(1)_Y^{}\times U(1)_X^{} \times U(1)_{PQ}\Longrightarrow U(1)_{em}^{}$, we can express the Higgs singlets and doublets to be 
\begin{eqnarray}
\label{vhp}
\sigma &=& \frac{1}{\sqrt{2}}\left(v_\sigma^{} + h_\sigma^{} + i P_\sigma^{}\right),~~\xi = \frac{1}{\sqrt{2}}\left(v_\xi^{} + h_\xi^{} + i P_\xi^{}\right),\nonumber\\
[2mm]
\phi&=&\left[\begin{array}{c}\phi_{}^{+}\\
[2mm]
\frac{1}{\sqrt{2}}\left(v_\phi^{}+ h_\phi^{} + i P_\phi^{}\right)\end{array}\right],\nonumber\\
[2mm]
\eta&=&\left[\begin{array}{c}\eta_{}^{+}\\
[2mm]
\frac{1}{\sqrt{2}}\left(v_\eta^{}+h_\eta^{} + i P_\eta^{}\right)\end{array}\right].
\end{eqnarray}
In addition to the four massive Higgs bosons $h_\xi^{}$, $h_\sigma^{}$, $h_\phi^{}$ and $h_\eta^{}$, i.e.
\begin{widetext}
\begin{eqnarray}
V&\supset& \lambda_\sigma^{} v_\sigma^2 h_\sigma^2  +\frac{1}{2} \left(2 \lambda_\xi^{} v_\xi^2 - \frac{\rho v_\phi^{} v_\eta^{}}{\sqrt{2}v_\xi^{}}\right)h_\xi^2 + \frac{1}{2}\left(2\lambda_\phi^{} v_\phi^2 - \frac{\rho v_\xi^{} v_\eta^{}}{\sqrt{2}v_\phi^{}}\right) h_\phi^2 + \frac{1}{2}\left(2\lambda_\eta^{} v_\eta^2 - \frac{\rho v_\xi^{} v_\phi^{}}{\sqrt{2}v_\eta^{}}\right)  h_\eta^2 +\lambda_{\sigma\xi}^{}v_\sigma^{} v_\xi^{} h_\sigma^{} h_\xi^{} \nonumber\\
&&+ \lambda_{\sigma\phi}^{}v_\sigma^{} v_\phi^{} h_\sigma^{} h_\phi^{} + \lambda_{\sigma\eta}^{}v_\sigma^{} v_\eta^{} h_\sigma^{} h_\eta^{}+ \left(\lambda_{\xi\phi}^{}v_\xi^{}v_\phi^{} + \frac{1}{\sqrt{2}}\rho v_\eta^{}\right)h_\xi^{} h_\phi^{} + \left(\lambda_{\xi\eta}^{}v_\xi^{}v_\eta^{} + \frac{1}{\sqrt{2}}\rho v_\phi^{}\right)h_\xi^{} h_\eta^{} \nonumber\\
&&+  \left[\left(\lambda_{\phi\eta}^{}+\lambda'^{}_{\phi\eta} \right)v_\phi^{}v_\eta^{} + \frac{1}{\sqrt{2}}\rho v_\xi^{}\right]h_\phi^{} h_\eta^{} \,, 
\end{eqnarray}
\end{widetext}
there should be five massless NG bosons, 
\begin{widetext}
\begin{eqnarray}
\label{ngb123}
G^{\pm}_{W}&=& \frac{v_\phi^{}\phi^{\pm}_{}+v_\eta^{}\eta^{\pm}_{}}{\sqrt{v_\phi^2+v_\eta^2}}\,,~~
G^{0}_{Z}= \frac{v_\phi^{}P^{}_{\phi} + v_\eta^{} P^{}_{\eta}}{\sqrt{v_\phi^2+v_\eta^2}}\,,~~
G^{0}_{X}=\frac{v_\phi^{} v_\eta^{2} P_\phi^{} - v_\phi^2 v_\eta^{} P_\eta^{} + v_\xi^{}\left(v_\phi^2+v_\eta^2\right) P_\xi^{} + v_\sigma^{} \left(v_\phi^{2}+v_\eta^2\right) P^{}_{\sigma}}{\sqrt{\left(v_\phi^2+v_\eta^2\right)\left[v_\sigma^2  \left(v_\phi^2 + v_\eta^2 \right)+v_\xi^2 \left(v_\phi^2+v_\eta^2\right)+  v_\phi^{2}v_\eta^2 \right]}}\,;\\
[2mm]
\label{axion}
a&=&\frac{- v_\sigma^{} \left[v_\phi^{} v_\eta^{2} P_\phi^{} - v_\phi^2 v_\eta^{} P_\eta^{} + v_\xi^{}\left(v_\phi^2+v_\eta^2\right) P_\xi^{} \right]+ \left[ v_\phi^2 v_\eta^2 +v_\xi^{2} \left(v_\phi^{2}+v_\eta^2\right) \right]P^{}_{\sigma}}{\sqrt{\left[v_\phi^2v_\eta^2+v_\xi^2\left(v_\phi^2+v_\eta^2\right)\right]\left[v_\sigma^2  \left(v_\phi^2 + v_\eta^2 \right)+v_\xi^2 \left(v_\phi^2+v_\eta^2\right)+   v_\phi^{2}v_\eta^2 \right]}}\,,
\end{eqnarray}
\end{widetext}
one massive pseudo scalar,
\begin{eqnarray}
\label{pscalar}
P&=&\frac{ - v_\xi^{}v_\eta^{} P_\phi^{}+ v_\xi^{} v_\phi^{} P_\eta^{} + v_\phi^{} v_\eta^{} P_\xi^{}}{ \sqrt{v_\phi^2 v_\eta^2 + v_\xi^{2}v_\phi^2 +v_\xi^{2} v_\eta^{2}}}~~\textrm{with}\nonumber\\
&&m_P^2 = - \frac{1}{\sqrt{2}} \rho v_\xi^{} v_\phi^{} v_\eta^{} \left(\frac{1}{v_\xi^{2}} + \frac{1}{v_\phi^{2} }+\frac{1}{v_\eta^{2} }\right),
 \end{eqnarray}
and one massive charged scalar,
 \begin{eqnarray}
 \label{charged}
H^{\pm}_{}&=& \frac{-v_\eta^{}\phi^{\pm}_{}+v_\phi^{}\eta^{\pm}_{}}{\sqrt{v_\phi^2+v_\eta^2}}~~\textrm{with}\nonumber\\
&&m_{H^\pm_{}}^2=-\left(\frac{1}{2}\lambda'^{}_{\phi\eta}+ \frac{\rho v_\xi^{}}{\sqrt{2}v_\phi^{} v_\eta^{}}\right)\left(v_\phi^2 + v_\eta^2 \right)\,.
\end{eqnarray}
While the four NG bosons $G^{\pm}_{W}$, $G^0_{Z}$ and $G^0_X$ are eaten by the longitudinal components of the SM gauge bosons $W^{\pm}_{}$ and $Z^0_{}$ as well as the $U(1)_X^{}$ gauge boson $X$, the fifth NG boson $a$ keeps available to be an axion. Actually, the two NG bosons $G^0_X$ and $a$ are the linear combinations of the two NG bosons from the $U(1)_t^{}$ and $U(1)_T^{}$ global symmetry breaking, i.e. the $P_\phi^{}$, $P_\eta^{}$ and $P_\xi^{}$ part in the $G^0_X$ and $a$ formula is according to the $U(1)_t^{}$ NG boson while the $P_\sigma^{}$ part is according to the $U(1)_T^{}$ NG boson.

We can deduce the coupling of the axion $a$ to the gluons $G^{a}_{\mu\nu}$ by 
\begin{eqnarray}
\mathcal{L} &\supset& - \frac{P_\sigma ^{}}{2v_\sigma^{}}  \partial_\mu^{}\left(\bar{T}\gamma^\mu_{}\gamma_5^{} T\right) - \frac{P_\phi^{}}{2v_\phi^{}}  \partial_\mu^{}\left(\bar{t}\gamma^\mu_{}\gamma_5^{} t\right)\nonumber\\
& =&\frac{a}{f_a^{}} \frac{\alpha_s^{}}{8\pi}G^{a}_{\mu\nu} \tilde{G}^{a\mu\nu}_{} \,,
\end{eqnarray}
where the axion decay constant is given by 
\begin{eqnarray}
\!\!\!\!f_a^{}&=&2v_\sigma^{} \sqrt{v_\sigma^2  \left(v_\phi^2 + v_\eta^2 \right)+v_\xi^2 \left(v_\phi^2+v_\eta^2\right)+   v_\phi^{2}v_\eta^2 }\nonumber\\
&&\times \frac{\sqrt{v_\phi^2v_\eta^2+v_\xi^2\left(v_\phi^2+v_\eta^2\right)}}{\left[  v_\sigma^2 v_\eta^2 - v_\xi^2\left(v_\phi^2 +  v_\eta^2\right)  - v_\phi^2 v_\eta^2 \right]}\,.
\end{eqnarray}
Remarkably, only the top quark $t$ coupling to the Higgs doublet $\phi$ and the new up-type quark $T$ coupling to the Higgs singlet $\sigma$ contribute to the axion-gluon coupling while the contributions from the other quarks coupling to the Higgs doublet $\eta$ have been exactly cancelled by each other. This result is consistent with our starting point: the right-handed top quark $t_R^{}$ and the left-handed new quark $T_L^{}$ are the unique fermions carrying the nontrivial PQ charges. Moreover, the top quark and the new quark are much heavier than the QCD scale so that the axion in the present model should obtain its mass like the KSVZ axion rather than the DFSZ axion.

\section{Symmetry breaking scales}

Now the bottom quark mass $m_b^{}= y_b^{} v_\eta^{} /\sqrt{2}$ and the perturbation requirement $y_b^{}< \sqrt{4\pi}$ puts a low limit on the VEV $v_\eta^{}$, i.e. \cite{pdg2020}
\begin{eqnarray}
v_{\eta}^{}=\frac{m_b^{}}{y_b^{}/\sqrt{2}} > \frac{4.18\,\textrm{GeV}}{\sqrt{4\pi}/\sqrt{2}}=1.18\,\textrm{GeV}\,,
\end{eqnarray}
and hence offers a large factor \cite{pdg2020}, 
\begin{eqnarray}
\frac{v_\phi^2}{v_{\eta}^{2}}<\frac{\left(246\,\textrm{GeV}\right)^2_{}-\left(1.18\,\textrm{GeV}\right)^2_{}}{\left(1.18\,\textrm{GeV}\right)^2_{}}= 4.35\times 10^4_{}\,. 
\end{eqnarray}
We then assume 
\begin{eqnarray}
v_{\eta}^{2}\ll v_{\phi,\xi}^2\ll v_\sigma^2\,,
\end{eqnarray} 
to simplify Eqs. (\ref{ngb123}) and (\ref{axion}), i.e. 
\begin{eqnarray}
G_W^\pm &\simeq & \phi^\pm_{}+ \frac{v_\eta^{}}{v_\phi^{}}\eta^\pm_{}\,, ~~G_Z^0\simeq P_\phi^{} + \frac{v_\eta^{}}{v_\phi^{}} P_\phi\,,\nonumber\\
G_X^0 &\simeq & \frac{v_\eta^2}{v_\sigma^{} v_\phi^{}}P_\phi^{} -\frac{v_\eta^{}}{v_\sigma^{}}P_\eta^{} +\frac{v_\xi^{}}{v_\sigma^{}}P_\xi^{} + P_\sigma^{}\,;\\
a&\simeq &  -\frac{v_\eta^2}{v_\xi^{} v_\phi^{}}P_\phi^{} + \frac{v_\eta^{}}{v_\xi^{}}P_\eta^{} -P_\xi^{} + \frac{v_\xi^{}}{v_\sigma^{}} P_\sigma^{}\,,
\end{eqnarray}
as well as Eqs. (\ref{pscalar}) and (\ref{charged}), i.e. 
\begin{eqnarray}
P&\simeq &\frac{v_\eta^{}}{v_\xi^{}}P_\xi^{} -\frac{v_\eta^{}}{v_\phi^{}}P_\phi^{} + P_\eta^{}~~\textrm{with}\nonumber\\
&&m_P^2 \simeq -\frac{ \sqrt{2} \rho v_\xi^{} v_\phi^{} }{v_\eta^{}}\simeq m_{h_\eta}^2 \,;\\
H^{\pm}_{}&\simeq &-\frac{v_\eta^{}}{v_\phi^{}} \phi^{\pm}_{}+ \eta^{\pm}_{}~~\textrm{with}\nonumber\\
&&m_{H^\pm_{}}^2\simeq  m_{h_\eta}^2- \frac{1}{2}\lambda'^{}_{\phi\eta} v_\phi^2 \,.
 \end{eqnarray}
As for the axion decay constant $f_a^{}$, it can be also given by a simple formula,
\begin{eqnarray}
f_a^{}\simeq 2 v_\xi^{} \left(\frac{v_\phi^2}{v_\eta^2}\right) \left/ \left(1-\frac{v_\phi^2/v_\eta^2}{v_\sigma^2/v_\xi^2}\right)\right..
\end{eqnarray}

In the above simplification, the Higgs singlet $\sigma$ dominates the $U(1)_X^{}$ gauge symmetry breaking while the Higgs singlet $\xi$ dominates the $U(1)_{PQ}^{}$ global symmetry breaking. In consequence, the PQ symmetry breaking scale should be determined by the VEV $v_\xi^{}$. To mach the experimental constraint on the axion decay constant, i.e. $f_a^{}\gtrsim 10^{9}_{}\,\textrm{GeV}$ \cite{pdg2020}, the PQ symmetry breaking scale $v_\xi^{}$ thus should have the low bound as follows,
\begin{eqnarray}
v_\xi^{}& = & 1.15 \times 10^{4}_{}\,\textrm{GeV} \left(\frac{f_a^{}}{10^9_{}\,\textrm{GeV}}\right)\nonumber\\
&&\textrm{for}~~\frac{v_\phi^2}{v_{\eta}^{2}}= 4.35\times 10^4_{} \ll \frac{v_\sigma^2}{v_{\xi}^{2}}\,.
\end{eqnarray}
A more appealing possibility is to consider the technically natural parameter choice as follows, 
\begin{eqnarray}
c=1-\frac{v_\phi^2/v_{\eta}^{2}}{v_\sigma^2/v_{\xi}^{2}}\simeq \mathcal{O}\left(0.001-0.1\right),
\end{eqnarray}
and hence break the PQ symmetry near the weak scale, 
\begin{eqnarray}
v_\xi^{}& \simeq & f_a^{} \left(\frac{c}{v_\phi^2/v_\eta^2 }\right)\nonumber\\
&=&115\,\textrm{GeV} \left(\frac{f_a^{}}{10^9_{}\,\textrm{GeV}}\right)\left(\frac{4.35\times 10^4_{}}{v_\phi^2/v_{\eta}^{2}}\right)\left(\frac{c}{0.01}\right).~~~~
\end{eqnarray}
Accordingly, the $U(1)_X^{}$ gauge symmetry breaking scale is determined to be 
\begin{eqnarray}
v_\sigma^{}& \simeq  &v_\xi^{} \sqrt{\frac{v_\phi^2}{v_{\eta}^{2}}} \nonumber\\
&=&2.40\times 10^4_{}\,\textrm{GeV} \left(\frac{f_a^{}}{10^9_{}\,\textrm{GeV}}\right)\sqrt{\frac{4.35\times 10^4_{}}{v_\phi^2/v_{\eta}^{2}}}\left(\frac{c}{0.01}\right).\nonumber\\
&&
\end{eqnarray}
This means the $U(1)_X^{}$ gauge boson which has the mass and kinetic mixing with the SM gauge bosons can be also near the weak scale for a proper $U(1)_X^{}$ gauge coupling.

\section{Summary and discussion}

In this work we have demonstrated a simple, economical and practical mechanism to efficiently lower the breaking scale of PQ symmetry. Our basic model contains one gauge boson, one up-type vector-like quark, two Higgs doublets and two Higgs singlets besides the SM fermions and gauge bosons. In particular, the right-handed top quark is the unique SM fermion carrying a nontrivial PQ charge. Such special role can be well motivated by the fact that the top quark is much heavier than the other SM fermions. Thanks to the SM and new gauge interactions, the Higgs doublet for the top quark mass generation as well as the Higgs singlet for the new quark mass generation can only contribute a tiny fraction to the axion. The axion decay constant can be enhanced by a huge ratio between the two Higgs doublet VEV squares. This ratio can reduce the PQ symmetry breaking scale by four orders of magnitude, saying from the order $10^9_{}\,\textrm{GeV}$ to the order $10^{4-5}_{}\,\textrm{GeV}$. If the PQ symmetry breaking scale is further expected at the weak scale, we can take the ratio between the two Higgs singlet VEV squares to be almost the ratio between the two Higgs doublet VEV squares. In our mechanism, two independent global symmetries connected by the new gauge interaction are the origin of the physical PQ global symmetry, meanwhile, everyone of these two global symmetries only acts on either the right-handed top quark or the left-handed new quark so that the PQ symmetry can avoid the domain wall problem.

In the usual invisible axion models, the related new particles should be of the order of the PQ symmetry breaking scale and hence they could not leave any signals to be verified experimentally. However, our mechanism can allow the PQ symmetry, which is now accompanied by the $U(1)_X^{}$ gauge symmetry, to be broken at low scales even near the weak scale. The new particles in our mechanism could be near the weak scale and hence they may be verified at colliders. Furthermore, our basic model can easily accommodate the seesaw mechanism for small neutrino mass. Specifically, the PQ symmetry can enforce the Higgs doublet $\eta$ rather than the Higgs doublet $\phi$ to take part in the seesaw mechanism. Since the Higgs doublet $\eta$ acquires a VEV of the GeV order, the low scale seesaw such as the TeV scale seesaw can become more motivated.

\textbf{Acknowledgement}: This work was supported in part by the Fundamental Research Funds for the Central Universities.


\begin{thebibliography}{99}


\bibitem{pq1977}
R.D. Peccei and H.R. Quinn, Phys. Rev. Lett. \textbf{38}, 1440 (1977).





\bibitem{adler1969}
S.L. Adler, Phys. Rev. \textbf{177}, 2426 (1969).


\bibitem{bj1969}
J.S. Bell and R. Jackiw, Nuovo Cim. \textbf{60A}, 47 (1969).
 
\bibitem{ab1969} 
S.L. Adler and W.A. Bardeen, Phys. Rev. \textbf{182}, 1517 (1969).


\bibitem{weinberg1978}
S. Weinberg, Phys. Rev. Lett. \textbf{40}, 223 (1978).


\bibitem{wilczek1978}
F. Wilczek, Phys. Rev. Lett. \textbf{40}, 279 (1978).

 
\bibitem{kg2010} 
J.E. Kim and G. Carosi, Rev. Mod. Phys. \textbf{82}, 557 (2010).

\bibitem{dgnv2020}
L. Di Luzio, M. Giannotti, E. Nardi, and L. Visinelli, Phys. Rep. \textbf{870}, 1 (2020).

 
\bibitem{pdg2020}
P.A. Zyla {\it et al.} (Particle Data Group), Prog. Theor. Exp. Phys. \textbf{2020}, 083C01 (2020) and 2021 update.

 
\bibitem{kim1979}
J.E. Kim, Phys. Rev. Lett. \textbf{43}, 103 (1979).


\bibitem{svz1980}
M.A. Shifman, A. Vainshtein, and V.I. Zakharov, Nucl. Phys. B \textbf{166}, 493 (1980).


\bibitem{dfs1981}
M. Dine, W. Fischler, and M. Srednicki, Phys. Lett. B \textbf{104}, 199 (1981).


\bibitem{zhitnitsky1980}
A. Zhitnitsky, Sov. J. Nucl. Phys. \textbf{31}, 260 (1980).



\bibitem{cv2016}
J.D. Clarke and R.R. Volkas, Phys. Rev. D \textbf{93}, 035001 (2016).


\bibitem{mrs2001}
E. Ma, M. Raidal, and U. Sarkar, Phys. Lett. B \textbf{504}, 296 (2001).


\bibitem{cky2014}
K. Choi, H. Kim, and S. Yun, Phys. Rev. D \textbf{90}, 023545 (2014).



\bibitem{kr2016}
D.E. Kaplan and R. Rattazzi, Phys. Rev. D \textbf{93}, 085007 (2016).

\bibitem{ci2016}
K. Choi and S.H. Im, JHEP \textbf{1601}, 149 (2016).


\bibitem{gu2021}
P.H. Gu, arXiv:2106.13010 [hep-ph].

\bibitem{sikivie1982}
P. Sikivie, Phys. Rev. Lett. \textbf{48}, 1156 (1982).

\bibitem{pwy1986}
R.D. Peccei, T.T. Wu, and T. Yanagida, Phys. Lett. B \textbf{172}, 435 (1986).

\bibitem{kw1986}
L.M. Krauss and F. Wilczek, Phys. Lett. B \textbf{173}, 189 (1986).




\bibitem{cfty2015}
C.W. Chiang, H. Fukuda, M. Takeuchi, and T.T. Yanagida, JHEP \textbf{1511}, 057 (2015)

\bibitem{cfty2017}
C.W. Chiang, H. Fukuda, M. Takeuchi, and T.T. Yanagida, Phys. Rev. D \textbf{97}, 035015 (2018).


\bibitem{ctty2018}
C.W. Chiang, M. Takeuchi, P.Y. Tseng, and T.T. Yanagida, Phys. Rev. D \textbf{98}, 095020 (2018).

\end{thebibliography}
\end{document}